\documentclass[aps,twocolumn,floats,prl,nofootinbib,superscriptaddress]{revtex4-1}

\usepackage[dvips]{graphicx} %
\usepackage{graphicx,amsmath,amsfonts,amssymb,slashed}
\usepackage{bbold,wasysym}
\usepackage{graphicx}

\usepackage[usenames,dvipsnames]{xcolor} 

\usepackage{soul}

\definecolor{RedWine}{rgb}{0.743,0,0}
\definecolor{RoyalBlue}{rgb}{0.25,.41,.88}

\setstcolor{Red}

\newcommand{\eV}{\ensuremath{\mathrm{eV}}}

\begin{document}

\title{Dark energy from the string axiverse}

\author{Marc Kamionkowski}
\affiliation{Department of Physics and Astronomy, Johns Hopkins
     University, 3400 N.\ Charles St., Baltimore, MD 21218, USA}
\author{Josef Pradler}
\affiliation{Institute of High Energy Physics, Austrian
     Academy of Sciences, Nikolsdorfergasse 18, 1050 Vienna,
     Austria}
\author{Devin G.\ E.\ Walker}
\affiliation{SLAC National Accelerator Laboratory, 2575 Sand Hill Road, Menlo Park, CA 94025, USA}

\begin{abstract}
String theories suggest the existence of a plethora of axion-like
fields with masses spread over a huge number of decades.  Here
we show that these ideas lend themselves to a model of
quintessence with no super-Planckian field excursions and in which
all dimensionless numbers are order unity.  The scenario
addresses the ``why now'' problem---i.e., why has
accelerated expansion begun only recently---by
suggesting that the onset of dark-energy domination occurs
randomly with a slowly decreasing probability per unit logarithmic
interval in cosmic time.  The standard axion potential requires
us to postulate a rapid decay of most of the axion fields that
do no become dark energy.  The need for these decays is averted,
though, with the introduction of a slightly modified axion
potential.  In either case, a Universe like ours arises in
roughly 1 in 100 universes.  The scenario may have a host of
observable consequences.
\end{abstract}

\maketitle

We still lack a well-established explanation for the origin of
the accelerated cosmic expansion observed in the Universe today
\cite{SNIa}.  The simplest guess, Einstein's cosmological
constant, works fine but requires a new fundamental parameter
with the unpalatable dimensionless amplitude of $10^{-120}$.
Quintessence circumvents this problem by suggesting that the
apparent cosmological constant is simply the vacuum energy
associated with the displacement of a scalar field from the
minimum of its potential \cite{quintessence}.  Still,
quintessence does not solve the ``why now'' problem; i.e., why
the Universe transitions from decelerated expansion to
accelerated expansion only fairly recently, after the Universe
has cooled 30 orders of magnitudes below the Planck temperature.
Ideas that involve alternative gravity or
large extra dimensions also generally require, ultimately, the
tuning of some parameter to be extremely small
\cite{reviews}.  Ideas based upon the string
landscape \cite{Bousso:2000xa} and/or anthropic arguments
\cite{Weinberg:1987dv} suggest that the value of
the cosmological constant in our Universe just happens to be the
one, of $\sim10^{120}$, that allows intelligent observers.

Here we show that ideas from string theory lend themselves
to a quintessence explanation for cosmic acceleration that
addresses the ``why now'' problem.  
It has long been understood~\cite{axionDE,Carroll:1998zi}
that an axion-like field provides a natural candidate for a
quintessence field, as the shift symmetry can protect the
extraordinary flatness required of the quintessence potential.
This solution, though, requires either that the axion decay
constant have a super-Planckian value, or that the initial axion
misalignment angle is extremely close to the value $\pi$ that
maximizes the potential, an option that is considered
fine-tuned, if not overlooked (but, as we will see, will be
essential in our scenario).  The hypothesis that quintessence
is an axion-like field also requires that the dark-energy
density today must still be put in by hand.

String theory may give rise to a ``string axiverse''
\cite{Svrcek:2006yi,Arvanitaki:2009fg}, a family of $O(100)$
axion-like fields with masses that span a huge number of decades.
In each decade of the cosmic expansion, one of
these axions becomes dynamical and has some small chance to drive an
accelerated expansion.  These chances are determined by the
initial value, assumed to be selected at random, of the axion
misalignment angle.  There is thus some chance that the Universe
will expand by $\sim30$ decades in scale factor before it
undergoes accelerated expansion.  As we will see, this
probability turns out to be $\sim1/100$ given the
distribution of axion masses and symmetry-breaking scales
suggested by the string axiverse \cite{Arvanitaki:2009fg}, 
We thus have an explanation for dark energy that involves no
parameters that differ from unity by more than an order of
magnitude.  Although there is still some element of chance or
anthropic selection required, a Universe that looks like ours
arises as a $\sim$one-in-100 occurrence, rather than a
one-in-$10^{120}$ event.  The model also specifies precisely the
set of initial conditions post-inflation as the set of randomly
chosen misalignment angles.

In the remainder of the paper we describe the scenario and
clarify the assumptions made.  As we will see, some additional
mechanism must be postulated to account for the effects of the
fields that do {\it not} become dark energy.  One possibility is
that the heaviest axions in the scenario, which would otherwise
contribute an unacceptably large density at the time of
big-bang-nucleosynthesis (BBN), undergo rapid decay.  Another
possibility that evade the overclosure problem involve a slight
modification to the usual axion potential.

We postulate a collection of axion fields each labeled by an
integer $a=1,2,3,\ldots$, and we define the misalignment angle
$\theta_a \equiv \phi_a/f_a$, where $f_a$ is the axion decay
constant for the $a$th field $\phi_a$.  The Lagrangian for this
axion field is then,
\begin{equation}
     \mathcal{L}_a = - \frac{f_{a}^2}{2}
     (\partial_{\mu}\theta_a)^2 - \Lambda_a^4 U(\theta_a) ,
     \quad U(\theta) = 1 - \cos{\theta},
\end{equation}
and the angle $\theta_a$ resides in the interval $\theta_a\in
[-\pi,\pi]$.  Here $2\Lambda_a^4$ is the maximum vacuum energy
associated with the $a$th axion field.

There are a variety of ways in which axion fields may be
populated in string theory.  Here, to be concrete, we specify a
particular realization in which the parameters for the $a$th
field are \cite{Arvanitaki:2009fg},%
\footnote{Note that if we had chosen our decay constant to be that,
  $f_a=\alpha M_P/S_a$, suggested in
  Ref.~\protect\cite{Arvanitaki:2009fg}, the misalignment angle
  required for a field to become dark energy would have been more
  stringent by several orders of magnitude, and the $\sim1/100$
  probability we arrive at below reduced accordingly.}
\begin{equation}
\label{eq:axiverseS}
  \Lambda_a^4 = \mu^4 e^{-S_a}, \qquad f_a = \alpha M_P,
\end{equation}
where $\alpha\lesssim1$ is an order-unity constant, in line with the
theoretical prejudice that $f_a$ should be close to some
fundamental scale (Planck or GUT) where global symmetries
are expected to be broken. The choice of $\alpha\lesssim1$ is also
compatible with the conjecture that gravity is the weakest
force~\cite{ArkaniHamed:2006dz}.  Since $f_a\lesssim M_P$, field
excursions in the model are never super-Planckian.  
For our numerical illustrations below, we will take
$\alpha=0.1$.  Here, $M_P=(8\pi
G)^{-1/2}=2.43\times10^{18}$~GeV is the reduced Planck mass, and
$\mu$ is a mass parameter related to the geometric mean of the
supersymmetry-breaking scale and the Planck mass, which we take to
be $\mu= \mu_{12}\, 10^{12}$~GeV, with $\mu_{12}$ a dimensionless
constant. Finally, $S_a$ is the action of the string instanton that
generates the axion potential.  We will take it to be $S_a=\beta a$
with $\beta$ a dimensionless constant of order unity.

The mass of the $a$th axion is
\begin{equation}
\label{eq:param}
     m_a = \frac{\Lambda_a^2}{f_a}  \simeq   H_0
     \frac{\mu_{12}^2}{\alpha_{0.1}} e^{-(\beta a-223.1)/2}  , 
\end{equation}
where $H_0 \simeq 10^{-33}\, \eV$ is the present-day Hubble rate.  In
this model, the distribution of axion masses is constant per
logarithmic interval in axion mass; i.e., $dN/d (\log_{10} m) \simeq
4.6/\beta$.  Thus, if we take $\beta = 9$, there is about one axion for
every two decades of axion mass.

Now consider the time evolution of these axion fields. The
equation of motion for each scalar field is
\begin{align}\label{eq:eom}
  \ddot \theta_a + 3 H \dot \theta_a +m_a^2  \sin \left( \theta_a
  \right) = 0 ,
\end{align}
where the dot denotes the derivative with respect to cosmic time $t$.
The Hubble parameter is determined from the Friedmann equation,
\begin{equation}
     H^2=\frac{1}{3M_p^2}\left[ \sum_a \rho_a +
     \rho_m R^{-3} + \rho_r R^{-4} \right],
\end{equation}
where $R(t)$ is the scale factor normalized to $R=1$ today, and
$\rho_m$ and $\rho_r$ are respectively the matter and radiation
energy density today.  The energy density in the $a$th axion
field is given by,
\begin{equation}
  \rho_a  = \frac12 f_a^2 \dot\theta_a^2 + \Lambda_a^4 U(\theta_a).
\label{eq:omgphi}
\end{equation}
The pressure $p_a$ of the $a$th field is given by the same
expression but with the sign of second term (the potential)
reversed, and the equation-of-state parameter is $w_a=p_a/\rho_a$.

The axion equation of motion is integrated from $t=0$ with an initial
field value $\theta_{a,I}$ and $\dot\theta_a(t=0)=0$.  We surmise that
each $\theta_{a,I}$ is selected at random from a uniform
distribution in the range $-\pi < \theta_{a,I} < \pi$.  At
sufficiently early times that $m_a \lesssim 2 H$ the axion field
is frozen because of Hubble friction and its vacuum energy is
negligible compared with the matter/radiation
density.  

There are then two possibilities for the subsequent
evolution after the Universe cools sufficiently so that $m_a \simeq 2H$,
when the scale factor is $\sim R_a$, determined by $2
H(R_a)\simeq m_a$.  The
first possibility, which occurs if $|\theta_{a,I}|$ is not too
close to $\pi$, is that the axion field begins to oscillate and behaves as
nonrelativistic matter with an energy density that decreases as
$\rho_a \sim \Lambda_a^4 [1-\cos(\theta_{a,I})] (R_a/R)^3$.  

The second possibility, which occurs if $|\theta_{a,I}|$ is close
enough to $\pi$, is that the axion field rolls slowly towards its
minimum, with an equation-of-state parameter $w_a < -1/3$. In this
case, the energy density of this axion field may come to dominate the
cosmic energy budget and drive a period of accelerated
expansion.  The condition for the field to roll slowly is
$\epsilon = (M_P^2/2)(V'/V)^2<1$, which requires the initial
misalignment angle to be in the range
\begin{equation}
   \pi > |\theta_{a,I}| \gtrsim \pi - 2\sqrt{2} f_a/M_p = \pi -
   2 \sqrt{2} \alpha.
\label{eqn:range}
\end{equation}
The probability that the $a$th field will drive accelerated
expansion is thus $\sim2\sqrt{2} \alpha /\pi \simeq \alpha$.

We now take an initial time near the onset of radiation domination,
after inflation and reheating, when the Universe has a temperature
$T_{\rm re}$ which we suppose is $T_{\rm re}\gtrsim\mu$.  We
suppose that at this time, the
initial field values $\theta_{a,I}$ for the fields that enter the
horizon after reheating (those with $m_a \lesssim T_{\rm re}$)
have been fixed during inflation.  The successful dark-energy
model that we seek is then one in which there is no accelerated
expansion from $T_{\rm re}$ until very recent times, redshift
$z\sim1$, at which point the Universe enters a period of
dark-energy domination.

Recall that the axion mass decreases monotonically (for $a\gtrsim$~few) with
$a$ and that the Hubble parameter decreases with time.  Therefore, the
different axion fields become dynamical ($m_a \simeq 2H$) in a sequence
of increasing $a$.  There is some small $\sim\alpha$ chance
that the first axion ($a=1$) would drive
accelerated expansion, and if so, that cannot describe our
Universe.  Suppose, though, that it does not drive accelerated
expansion.  There is then another $\sim\alpha$ chance that the
second axion field will drive
accelerated expansion.  If it does, then that is not our
Universe.  Cumulatively, the chance that the first $a-1$ fields
do not drive accelerated expansion but that $a$th field does is
\begin{equation}
     P(a) = \alpha \left(1 - \alpha\right)^{a-1}.
\label{eqn:probability}
\end{equation}
This equation encapsulates
the heart of this model for dark energy.  The important point
is that there is for the relevant values of $a$ a slowly falling
probability per unit logarithmic interval in axion mass, for a
given axion to act as dark energy.  From this it follows that
{\it there is a slowly decreasing probability per logarithmic
interval in cosmic time (or scale factor or redshift or cosmic
temperature) for the Universe to become dark-energy dominated}.

The axion field that describes cosmic acceleration in our
Universe is one which has a density $2 m_a^2 f_a^2 \simeq
\Omega_\Lambda \rho_c\simeq 0.7\, (3\, H_0^2 M_P^2)$, where
$\Omega_\Lambda$ is the fraction of the critical density
$\rho_c =3\, H_0^2 M_p^2$ in dark energy today.  From this it
follows that the axion field responsible for cosmic acceleration
must have $m_a \simeq H_0/\alpha$.  Combining this with
Eq.~(\ref{eq:param}), we find that if $\mu_{12}=1$,
$\alpha=0.1$, and $\beta = 9$, for example, then the field that
becomes dark energy has index $a \simeq 24$ (and $m_a = 2
H(R_a)$ is met at redshift $z \simeq 2$).  From
Eq.~(\ref{eqn:probability}), the probability that a
given Universe will have a 
cosmological constant like that we observe, with $a=24$
is then $\sim 1/100$.  {\it We conclude that if we were to look at
100 post-inflation universes with different randomly selected sets of initial
field values $\theta_{a,I}$, we would expect one of them to wind
up looking like our Universe}.  We thus have an explanation for
cosmic acceleration that invokes no dimensionless parameters
that differ from unity by more than one order of magnitude.  The
results depend only logarithmically on $\mu$ and do not differ
considerably for order-unity changes to $\alpha$ and $\beta$.

The next step in the consideration of these models is to
understand the effects of the $\sim 23$ axion fields that do not
become dark energy.  A typical such field will have
$\theta_{a,I}\sim1$ and will, when it begins to oscillate when
$m_a\sim H$, have an energy density $\sim m_a^2
f_a^2$ that is smaller than the critical density $\sim H^2
M_P^2$ by a factor $\sim \alpha^2$.  However, the energy density
in the coherent field oscillations that ensue scales with the
scale factor as $R^{-3}$, as opposed to the radiation density,
which scales as $R^{-4}$.  The energy density of these fields,
especially those that enter the horizon first, thus comes to
dominate the energy density of the Universe.  If these axions do
not decay, they overclose the Universe by a huge amount; for
example, the first field ($a=1$), with $m_a\sim 10^3$~GeV, would
have an energy density today $\sim 10^{11}$ times the current
critical density.

If, however, the axions with masses $m_a \gtrsim H_{\rm bbn} \sim
10^{-17}$~eV---those with $a<a_{\rm bbn}\simeq 16$---decay on a
timescale less than $\sim$sec (when BBN begins) to Standard Model
particles, then they will simply heat the Universe, without any
observational consequences (apart from the dilution of any
pre-existing dark-matter density or baryon number).  Such decays,
though, are unlikely to be sufficiently rapid, for axions at the
lower-mass end of this mass range, for a parametric decay rate $\Gamma
\sim m_a^3/f_a^2$, or even $\Gamma\sim m_a$.  Another possibility,
though, is that only a handful of the heaviest axions decay prior to
BBN.  If the $a$th axion did not decay, it would contribute, following
the reasoning above, a density $(\rho_a/\rho_R)_{\rm bbn} \propto
e^{-S_a/4}$.  From this it follows that the undecayed-axion energy
density at BBN is dominated by only a handful of the heaviest axions,
those with the smallest values of $a$.  With reasonable shifts to the
values of the parameters we have chosen (e.g., a larger $\mu$), decay
rates greater $\Gamma \gtrsim H_{\rm bbn}$ are not implausible for
fields with $a\sim$~few.  If these heaviest axions then decay to SM
particles before BBN, they can dilute the contributions of the lighter
axion fields to the cosmic energy budget at BBN.

Next consider the axions that enter the horizon after BBN but before
CMB decoupling ($a_{\rm cmb} \simeq 22$).  The energy density of these
axions must not exceed the bound on the dark-matter density at CMB
decoupling inferred from the CMB.  Again, the problem is dominated by
the most massive axion, that with $m_a\sim 10^{-17}$~eV, with $a\simeq
16$, that enters the horizon near the time of BBN.  The contribution
of this axion to the critical density at that time is $\sim
\alpha^{-2}\simeq10^{-2}$.  Since its density scales as $R^{-3}$, its
density, if it did not decay, would be $\Omega_a\sim 20$ today.  This,
and the few fields that enter the horizon post-inflation must decay
effectively to radiation. The axion field that enters the horizon at
redshift $z\sim 40$, with $a\simeq 23$ and $m_a\sim 10^{-30}$~eV would
naturally have the correct dark matter density. However, structure
formation requires that the fields preceding decoupling constitute the
bulk of the matter budget.

Some thought should be given to isocurvature perturbations, since the
axions that decay or that may make up the dark matter (or part of it)
have fluctuations that are not correlated with the curvature
perturbations induced during inflation.  However, if the primordial
plasma of SM particles is due primarily to the decays of the heaviest
axions, then the resulting perturbations are likely to be mostly
adiabatic.  If the dark matter turns out to be axions, then there may
be some worry that the perturbations in the axion--dark-matter density
may be isocurvature.  However, the scenario does not require these
axions to make up the dark matter. Even if they are the dark matter,
there may be mechanisms (e.g., Ref.~\cite{Higaki:2014ooa}) to avoid
problems with isocurvature perturbations.  If the scenario proceeds
via the $(1-\cos\theta)^3$ potential discussed below, then there are
no isocurvature perturbations. In summary, it will be important to
insure that isocurvature perturbations are not a problem in any
detailed implementation of the scenario that we outline, but
isocurvature perturbations are not necessarily a showstopper.

The principal objection one might have to the scenario above is
the rapid decays required of many of the axion fields that do not
become dark energy.  We now propose a slightly revised
scenario in which there are no problems with overclosure, and no
field decays required.

Suppose that the axion potential function $U(\theta)
=1-\cos\theta$ is replaced by $U(\theta)= (1-\cos\theta)^3$.
The broad outline of the scenario described above remains the
same.  The principal difference, though, is that once the field
begins to oscillate, it oscillates about a minimum that is
$V(\phi) \propto \phi^6$, rather than $\phi^2$.  Such
oscillations behave as matter with equation-of-state parameter
$w=1/2$ and have an energy density that decays with scale factor as
$R^{-9/2}$ \cite{Turner:1983he,Johnson:2008se}, more rapidly
than radiation, which decays as $R^{-4}$.  Thus, the energy
density in the fields that do not become dark energy always
remains negligible compared with the dominant radiation and
matter densities.

The slow-roll condition $\epsilon = (M_p^2/2)(V'/V)^2<1$ for
this altered potential is $\pi > \theta_{a,I} > \pi-(2\sqrt{2}/3)
\alpha$, more restrictive than in the first scenario
(for the same $\alpha$ and $\beta$) by a factor of three.  The
probability for the $a$th field to have $w<-1/3$ is obtained
from Eq.~(\ref{eqn:probability}) with the replacement
$\alpha \to \alpha/3$.

Although there is no mass associated with the field now (the
curvature about the minimum of the potential is zero), there is
an oscillation frequency $\omega_a(\phi_0)$ that depends on the
amplitude $\phi_0$ of the oscillation.  This frequency is small
as the misalignment-angle amplitude $\theta_0=\phi_0/f_a \to
\pi$ and then increases as $\theta_0$ decreases to $\theta_0
\sim 2.38$ at which point the oscillation frequency is $\omega_a
\simeq 0.69\, \Lambda_a^2/f_a$.  The subsequent decrease
$\omega_a \propto R^{-3/2}$ of the oscillation frequency is
slower than the decrease $H\propto R^{-2}$ of the Hubble
parameter.  Thus, once the Hubble parameter has decreased below
$\omega_a$, and the field begins to oscillate, it continues to
oscillate thereafter.  The energy density thus becomes
negligible compared with radiation, as claimed above.

We now recapitulate and then make closing remarks.  We suppose that
there are several hundred axion fields with masses and decay constants
distributed as in Eqs.~(\ref{eq:axiverseS}) and (\ref{eq:param}).
With the mass parameter $\mu$ chosen to be $\sim 10^{12}$~GeV, the
highest-mass axion (that with $a=1$) has a mass $\sim
  10^{3}-10^{4}$~GeV, and the axion with mass $m_a\sim H_0\sim
10^{-33}$~eV comparable to the Hubble parameter today has $S_a\simeq
219$, or $a \simeq 24$ if $\beta=9.0$.  We then surmise that after
inflation, which presumably reheats the Universe to a temperature
$T\gtrsim \mu$, the initial misalignment angles $\theta_{a,I}$ for all
the fields are selected at random.  We then argue that there is, with
the canonical parameters chosen, a roughly 1 in 100 chance that the
Universe will undergo radiation- and matter-dominated expansion phases
until a redshift $z\sim1$ when the dark energy associated with the
field with $a\simeq24$ will take over.  Our Universe turns out to be
this one-in-100 occurrence either because of the luck of the draw
and/or from some anthropic selection.  In this regard, the scenario
may require anthropic elements similar to those in landscape
scenarios.  The one-in-100 coincidence, though, does not make as
stringent demands on the imagination as a one-in-$10^{120}$
coincidence.

The $\sim 0.01$ probability we arrive at depends only weakly to
order-unity variations in the parameters $\alpha$ and $\beta$ and only
logarithmically on $\mu$.  This probability,
Eq.~(\ref{eqn:probability}), is obtained by requiring that none of the
axion fields that enter the horizon before that (with $a=24$) that
gives rise to the observed accelerated expansion lead to accelerated
expansion.  This requirement, though, may be too restrictive as some
of these earlier dark-energy-dominated phases may be relatively short
and have little observable impact.  If so, then the probability that a
Universe like the one we observe arises may be a bit larger, closer to
$\sim 1/10$.  Conversely, though, measurements that constrain the
equation-of-state parameter $w$ to be relatively close to $-1$ suggest
that the $a=24$th field that drives the observed dark energy must have
$\epsilon$ small compared with unity, and thus an initial misalignment
angle closer to $\pi$ than the full range allowed by
Eq.~(\ref{eqn:range}).  This will then decrease the probability
accordingly.  It may also be interesting to explore how the scenario
is modified if fields are populated with a distribution different from
that, in which $S_a$ is populated uniformly in $a$, assumed here.  The
scenario requires that we postulate either a rapid decay of most of
the fields that do not become dark energy, or a non-standard axion
potential in which the effects of these other fields quickly redshift
away.  It will be interesting to consider more complete models in
which these requirements arise~\cite{Pradler:inprep}.

There may also be, depending on the detailed implementation of the
ideas presented here, observational consequences of this scenario.
First of all, the model predicts a quintessence-dominated Universe
today, with a value of the observed equation-of-state parameter $w$
that differs from $-1$---i.e., the dark energy is not a cosmological
constant;\footnote{The scenario we present is thus distinct from that
  envisionsed in Ref.~\cite{Svrcek:2006hf}, that is similarly in
  spirit to N-flation \cite{Dimopoulos:2005ac}, in which there is a
  cosmological constant that is due to the displacement of the many
  axion fields with $m_a < H_0$ from their minimum.} this should be
tested with forthcoming measurements of the expansion history
\cite{searchtechniques}.  In this scenario, cosmic acceleration is due
to a quintessence field, and {\it not} a modification to gravity;
cosmic modified-gravity searches \cite{Joyce:2014kja} should therefore
all turn up null results.  If the field has an axion-like coupling to
electromagnetism, there may be cosmic birefringence
observed~\cite{Carroll:1998zi,Lue:1998mq}.  Residual decays of the
fields that enter the horizon before BBN or shortly after BBN may give
rise to spectral distortions in the CMB (see, e.g.,
Ref.~\cite{Chluba:2014sma} and references therein) or the spectrum of
primordial perturbations on very small scales \cite{Jeong:2014gna}.
There may be some component of primordial perturbations that is
isocurvature.  There may be observable consequences for astrophysical
black holes \cite{Arvanitaki:2010sy}.  And the dark matter may be
composed, at least in part, of axions, something that may be sought
with future experiments \cite{Graham:2013gfa}.  It will be interesting
to study these possibilities in more detailed implementations of the
ideas presented here.

\smallskip We thank A.~Arvanitaki for useful discussions.  MK thanks
the Aspen Center for Physics, supported by NSF Grant No.\ PHY-1066293,
for hospitality.  MK acknowledges the support at JHU of NSF Grant No.\
0244990, the John Templeton Foundation, and the Simons Foundation.  JP
is supported by the New Frontiers program of the Austrian Academy of
Sciences. DW is supported in part by a grant from the Ford Foundation
via the National Academies of the Sciences as well as the US
Department of Energy, contract DE-ACO2-76SF00515.

\end{document}